\documentclass[preprint,sor4-1,longbibliography]{revtex4-1}
\usepackage{mathtools}
\usepackage[table]{xcolor}
\definecolor{lightgray}{gray}{0.9}
\usepackage{graphics}
\usepackage{graphicx}

\begin{document}

\title{Direct observation of DNA dynamics in semi-dilute solutions in extensional flow}

\author{Kai-Wen Hsiao}
\affiliation{Department of Chemical and Biomolecular Engineering, University of Illinois at Urbana-Champaign, 600 S. Mathews Avenue, Urbana, IL 61801, USA}
\author{Chandi Sasmal}
\author{J. Ravi Prakash}
\affiliation{Department of Chemical Engineering, Monash University, 20 Research Way, Melbourne, VIC 3800, Australia}
\author{Charles M. Schroeder}
\email{cms@illinois.edu}
\affiliation{Department of Chemical and Biomolecular Engineering, University of Illinois at Urbana-Champaign, 600 S. Mathews Avenue, Urbana, IL 61801, USA}
\affiliation{Department of Materials Science and Engineering, University of Illinois at Urbana-Champaign, Urbana, IL 61801}

\date{\today}

\begin{abstract}
The dynamic behavior of semi-dilute polymer solutions is governed by an interplay between solvent quality, concentration, molecular weight, and flow type. Semi-dilute solutions are characterized by large fluctuations in polymer concentration, wherein polymer coils interpenetrate but may not be topologically entangled at equilibrium. In non-equilibrium flows, it is generally thought that polymer chains can `self-entangle' in semi-dilute solutions, thereby leading to entanglements in solutions that are nominally unentangled at equilibrium. Despite recent progress in the field, we still lack a complete molecular-level understanding of the dynamics of polymer chains in semi-dilute solutions. In this work, we use single molecule techniques to investigate the dynamics of dilute and semi-dilute solutions of $\lambda$-phage DNA in planar extensional flow, including polymer relaxation from high stretch, transient stretching dynamics in step-strain experiments, and steady-state stretching in flow. Our results are consistent with a power-law scaling of the longest polymer relaxation time $\tau \sim (c/c^{\ast})^{0.48}$ in semi-dilute solutions, where $c$ is polymer concentration and $c^{\ast}$ is the overlap concentration. Based on these results, an effective excluded volume exponent $\nu$ $\approx$ 0.56 was found, which is in good agreement with recent bulk rheological experiments. We further studied the non-equilibrium stretching dynamics of semi-dilute polymer solutions, including transient (1 $c^{\ast}$) and steady-state (0.2 $c^{\ast}$ and 1 $c^{\ast}$) stretching dynamics in planar extensional flow using an automated microfluidic trap. Our results show that polymer stretching dynamics in semi-dilute solutions is a strong function of concentration. In particular, a decrease in transient polymer stretch in semi-dilute solutions at moderate Weissenberg number ($Wi$) compared to dilute solutions is observed. Moreover, our experiments reveal a milder coil-to-stretch transition for semi-dilute polymer solutions at 0.2 $c^{\ast}$ and 1 $c^{\ast}$ compared to dilute solutions. Interestingly, a unique set of molecular conformations during the transient stretching process for single polymers in semi-dilute solutions is observed, which suggests transient stretching pathways for polymer chains in semi-dilute solutions are qualitatively different compared to dilute solutions due to intermolecular interactions. Taken together, this work provides a molecular framework for understanding the non-equilibrium stretching dynamics of semi-dilute solutions in strong flows. 
\end{abstract}

\maketitle

\section{Introduction}
The dynamics of semi-dilute polymer solutions is an intriguing yet particularly challenging problem in soft materials and rheology. Dilute polymer solutions are characterized by the rarity of overlap of single chains, whereas concentrated solutions and melts are governed by topological entanglements and dense polymer phases. Unentangled semi-dilute solutions, however, are characterized by coil-coil interpenetration at equilibrium, albeit in the absence of intermolecular entanglements under quiescent conditions. From this view, the dynamics of dilute solutions and concentrated solutions and melts can often be treated by the single chain problem or the framework of mean-field theories, which reduces the problem of many-body interactions in entangled solutions to the motion of a single polymer chain in an effective potential or field. On the other hand, semi-dilute polymer solutions are known to exhibit large fluctuations in concentration, which precludes the straightforward treatment of polymer dynamics in these solutions using a mean-field approach. 

\begin{figure}[t]
\includegraphics[height=10cm]{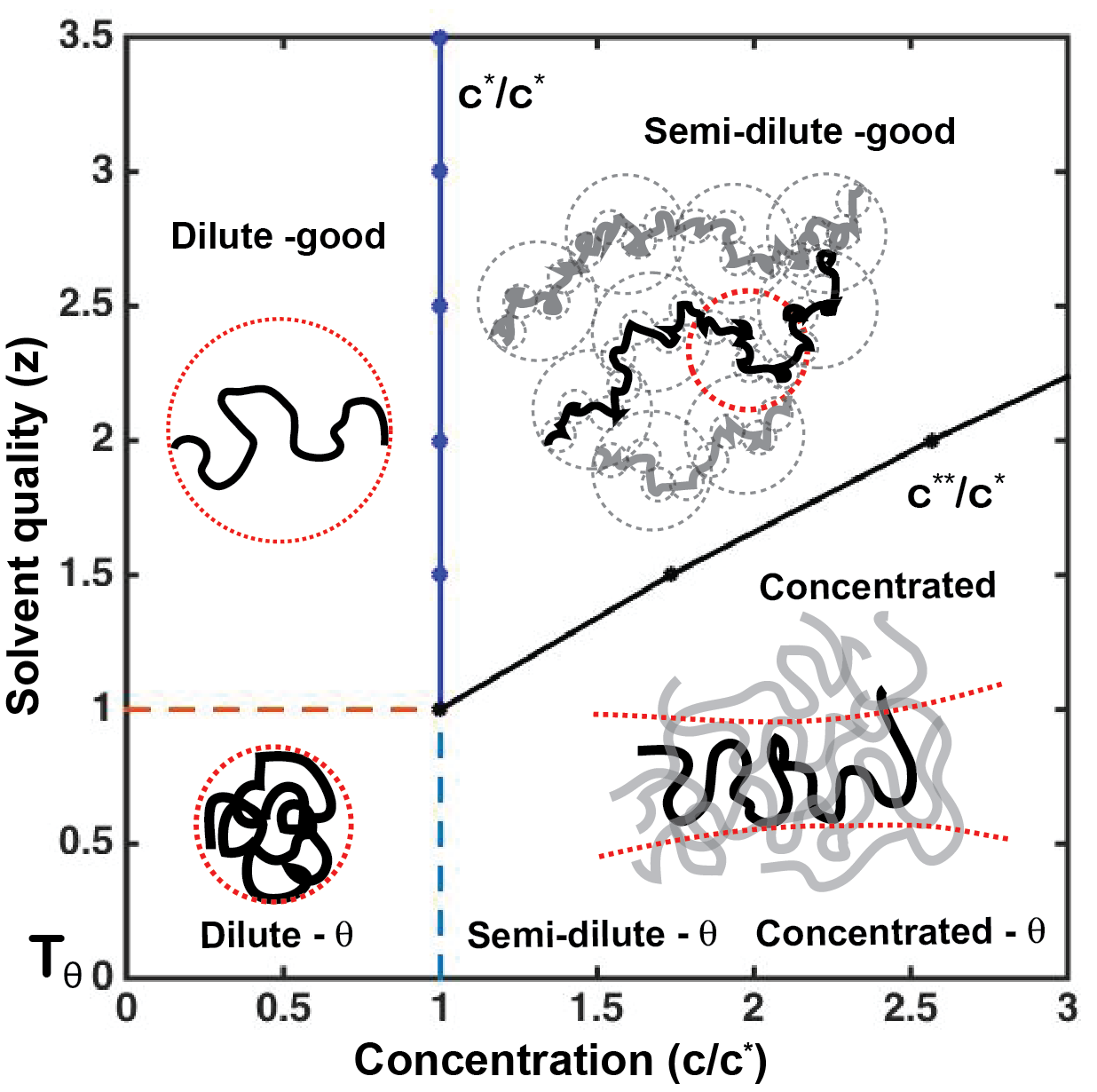}
  \caption{\label{fig:fig1}Phase diagram for polymer solutions as a function of relative concentration $c / c^{\ast}$ and solvent quality $z$ (see Section II for details). For display purposes, we chose monomer size $b$ = 1 and an excluded volume exponent $\nu = 0.56$ based on experimental results.}  
\end{figure}

The near equilibrium properties of semi-dilute polymer solutions are governed by an interplay between polymer concentration and solvent quality FIG.~\ref{fig:fig1} \cite{Jain2012,Rubinstein2003}. Two parameters are commonly used to describe the equilibrium properties of semi-dilute solutions. First, the critical overlap concentration $c^{\ast} \approx M / N_A R_g^3$ is used as a characteristic polymer concentration in semi-dilute solutions, where $M$ is polymer molecular weight, $N_A$ is Avogadro's number, and $R_g$ is the radius of gyration \cite{Graessley1980}. Using the overlap concentration, a scaled polymer concentration of $c/c^{\ast}=1$ corresponds to a bulk solution concentration of polymer that is roughly equivalent to the concentration of monomer within a polymer coil of size $R_g$. In addition, solvent quality can be characterized by the solvent quality parameter $z$, which is a function of polymer molecular weight $M$ and temperature $T$ relative to the theta temperature $T_{\theta}$ (Section II).

The equilibrium properties of semi-dilute polymer solutions have been widely studied using bulk techniques such as dynamic light scattering \cite{Brown1986c, Li2008c, Zettl2009c}, where polymer diffusion and relaxation dynamics were reported for synthetic polymers and compared with blob theory. Upon increasing polymer concentration above the dilute limit, two distinct relaxation modes are observed in semi-dilute polymer solutions, with the longer time scale attributed to cooperative, segment-segment interactions between polymers. Bulk shear rheology has also been used to study semi-dilute solutions of synthetic polymers \cite{Takahashi1985d, Takahashi1988, Raspaud1995d}, where a scaling relation between zero-shear viscosity and concentration in the semi-dilute regime was found to depend on polymer type and solvent quality. 

Moving beyond equilibrium, the non-linear dynamics of semi-dilute polymer solutions in shear flow has been extensively studied using a combination of bulk rheological and rheo-optical measurements, including transient and steady shear rheology. In startup of shear flow, a stress overshoot is observed in semi-dilute polymer solutions \cite{Chow1985, Magda1993, Kume1997}, which is attributed to the transient molecular stretching cycle of polymers in shear flow. The dynamics of semi-dilute solutions in extensional flows has also been studied using bulk rheological techniques. Extensional flow generally consists of an axis of fluid compression and an orthogonal axis of extension in the absence of fluid rotation. For this reason, extensional flows are considered as ``strong flows'' capable of stretching polymers to high degrees of extension. In ultra-dilute polymer solutions, it is well known that long linear polymers undergo a coil-stretch transition in steady extensional flows \cite{Fuller1981}. The coil-stretch transition has also been studied in semi-dilute polymer solutions using these techniques \cite{Ng1993}. Bulk measurements based on flow-induced birefringence in extensional flow have revealed rich information about conformational orientation and anisotropy under controlled flow conditions and varying time scales \cite{Ng1993, Dunlap1987}. In these studies, a strong increase in stress and an inhibition of development of high strain rates for a nominally dilute polymer solution ($\sim$ 0.1 $c^*$) is observed in extensional flow. Upon increasing polymer concentration, a dilatant effect is observed due to the formation of transient networks in semi-dilute polymer solutions \cite{Odell1989}. In capillary thinning experiments, Clasen and coworkers \cite{Clasen2006} found that the longest relaxation times of monodisperse polystyrene solutions at moderate concentration (0.01 $\leq$ ${c}/{c^*}$ $\leq$ 1) rise substantially higher than the relaxation times extracted from small amplitude oscillatory shear (SAOS) experiments. Taken together, these results suggest that an increase in polymer concentration results in a larger impact on dynamics in extensional flows compared to shear flow. 

Strong flow modification and coupling between semi-dilute polymer solutions and extensional flow fields were also reported using flow birefringence experiments, bulk rheology measurements, and Brownian dynamics (BD) simulations. Using a four-roll-mill apparatus, Ng and Leal observed that flow birefringence decreased in semi-dilute polymer solutions relative to dilute solutions \cite{Ng1993}, which corresponds to a decrease in polymer stretch in semi-dilute solution flows. Chow and coworkers used an opposing jets apparatus to study semi-dilute polystyrene solutions in extensional flow and reported the development of pipe-like birefringent structure as the strain rate increased \cite{Chow1988}. Interestingly, velocimetry measurements showed that this structure is caused by a reduction in strain rate in the center of the flow field \cite{Chow1988}. Using an extensional rheometer, Sridhar and coworkers \cite{Tirtaatmadja1993,James1995} found that the transient viscosity of dilute and semi-dilute polyisobutylene solutions are an order of magnitude smaller than predicted by Batchelor's expression for viscosity of a suspension of elongated particles \cite{Batchelor1971}. Brownian dynamics simulations by Harrison et al. \cite{Harrison1999} and Stoltz et al. \cite{Stoltz2006} also revealed a decrease in the maximum attainable polymer deformation when results are scaled with a concentration dependent Weissenberg number $Wi_c = \tau_c \dot{\varepsilon}$, where $\tau_c$ is the longest relaxation time in dilute or semi-dilute solution conditions. Overall, these results suggest that flow-induced entanglements or interchain interactions may inhibit polymer chains from stretching to full extension in strong flows. 

In recent years, single molecule techniques have provided the ability to directly visualize the motion of single polymer chains, thereby revealing molecular-level information on distributions in polymer conformation that is generally obscured in bulk experiments. High molecular weight, double stranded DNA molecules have been used as model polymers for single molecule imaging, and the dynamic properties of DNA have been characterized using bulk and single molecule methods \cite{Shaqfeh2005c}, including dynamic light scattering \cite{Chirico1989, Langowski1987}, zero-shear viscosity studies \cite{Heo2005e, Pan2014}, and single molecule diffusion measurements \cite{Smith1996}. Recently, Prakash and coworkers characterized the behavior of dilute and semi-dilute DNA solutions across a wide-range of solvent qualities from theta solvents to good solvents as a function of polymer molecular weight $M$ \cite{Pan2014}. These authors used dynamic and static light scattering to measure the hydrodynamic radius $R_H$ and theta temperature $T_{\theta}$ for DNA solutions, thereby enabling determination of $R_g$ and $c^{\ast}$ as a function of DNA molecular weight and temperature. In this way, this work provided a systematic framework to understand the concentration and temperature dependence of DNA-based polymer solutions. Furthermore, this work also elucidated the dynamic double crossover behavior in scaling for semi-dilute polymer solutions, wherein polymer behavior is considered in the context of smooth crossover regimes in solvent quality between theta and athermal solvents \cite{Pan2014}. In the last decade, several mesoscopic simulation techniques have been developed to study the non-equilibrium flow behavior of semi-dilute solutions \cite{Stoltz2006, Huang2010d, Jain2012a, Saadat2015}. In highly non-equilibrium flows, the screening of excluded volume (EV) interactions and intra- and intermolecular hydrodynamic interactions (HI) across multiple length scales is thought to play a major role on dynamics in non-equilibrium flows, and these effects can now be articulated by capitalizing on the aforementioned prior work.

Single polymer techniques have also been used to study the dynamics of single DNA molecules under highly non-equilibrium flow conditions in different flow types \cite{Shaqfeh2005c}. Using $\lambda$-phage DNA, Chu and co-workers studied the dynamics of single DNA polymers in ultra-dilute solutions (10$^{-5}$ c$^{\ast}$) in shear flow and planar extensional flow \cite{Perkins1994, Perkins1995, Perkins1997, Smith1999, Kantsler2012}. In the startup of extensional flow, it was found that identical polymers pass through different transient conformations under identical flow conditions due to subtle differences in their initial conformations and a delicate balance between convection and diffusion, a phenomenon known as ``molecular individualism'' \cite{Smith1998}. The coil-stretch transition in dilute solutions has also been studied for long linear polymers using single molecule imaging, where polymer conformation hysteresis is observed due to conformation-dependent intramolecular hydrodynamic interactions \cite{Schroeder2003, Schroeder2004}. Interestingly, molecular individualism was also observed in shear flow in both semi-dilute DNA solutions \cite{Hur2001, Babcock2000} and entangled DNA solutions \cite{Teixeira2007}, albeit with different molecular conformations compared to extensional flow. In the startup of shear flow, Hur et al.\cite{Hur2001} and Babcock et al.\cite{Babcock2000} observed a stress overshoot in semi-dilute DNA solutions (6 $c^*$) that was directly linked to polymer stretching conformations using single molecule imaging. More recently, Harasim et al. \cite{Harasim2013a} and Huber et al. \cite{Huber2014} directly observed the motion of semi-flexible actin filaments in semi-dilute solutions in shear flow using single molecule imaging and found significantly inhibited tumbling in shear. The `slowing down' of tumbling motion was attributed to the formation of transient structures due to intermolecular interactions.  

Despite recent progress in bulk rheology and single molecule studies, however, we still lack a complete understanding of the dynamics of semi-dilute polymer solutions in extensional flow. Given the importance and practical relevance of semi-dilute polymer solutions, it is crucial to develop a molecular-level picture of how polymers stretch and relax in semi-dilute polymer solutions. In this work, we use molecular rheology and single molecule imaging to explore the effect of polymer concentration on the transient and steady state dynamics of polymers in semi-dilute solutions in planar extensional flow. In particular, this work extends beyond prior single polymer studies in semi-dilute or concentrated solutions that have focused primarily on chain dynamics in shear flow. From this view, we aim to extend our understanding of single chain polymer dynamics in strong flows where intermolecular interactions play a key role in flow dynamics. In order to provide a comprehensive understanding of polymer dynamics in semi-dilute solutions, this paper is accompanied by a companion article describing Brownian dynamics simulations of single polymers in extensional flow in semi-dilute solutions \cite{Sasmal2016}, thereby directly complementing the experiments described in this article.

This paper is organized as follows. In Section II, polymer scaling theory in semi-dilute solutions in the context of the blob model is discussed. In Section III, we report experimental methods, including sample preparation of spatially homogeneous semi-dilute DNA solutions and optical imaging techniques. In Section IV, we characterize the longest relaxation times of single polymers in semi-dilute solutions, and further discuss these results in the context of bulk rheology data and theoretical predictions for semi-flexible polymers. Transient and steady state dynamics of single polymers in semi-dilute polymer solutions are characterized. Interestingly, a new set of molecular stretching conformations and pathways in startup of extensional flow in semi-dilute solutions is reported. We also discuss the steady-state stretching of polymers in semi-dilute extensional flows, where a milder coil-stretch transition compared to dilute solutions is observed. Finally, in Section V, our main findings are summarized with a brief conclusion.       

\section{Scaling theory and blob model} 
In dilute solutions, the near-equilibrium properties of polymer chains are determined by polymer molecular weight and solvent quality \cite{Rubinstein2003}. In theta conditions, a polymer chain can be described by an ideal random walk with root-mean-square end-to-end distance $R_0$. In good solvents, polymer chains tend to swell due to dominant intramolecular excluded volume (EV) interactions to yield an average coil size known as the Flory radius $R_{F}$, which is defined as the size of a real chain in the presence of EV interactions. From this view, the swelling ratio $\alpha_g \equiv \frac{R_F}{R_0}$ is a reflection of the solvent quality, with $\alpha_g >1$ corresponding to good solvent conditions. Solvent quality can be defined by the solvent quality parameter:
\begin{equation} 
z \equiv \left( \frac{3}{2 \pi} \right)^{3/2} \frac{\text{v}}{b^3} N^{1/2} = \left(\frac{3}{2\pi}\right)^{3/2} \left(1-\frac{T_{\theta}}{T}\right){N}^{1/2} \approx \frac{\text{v}}{b^3} N^{1/2} 
\end{equation}
where $\text{v}$ is the excluded volume of a real polymer chain, $N$ is the number of Kuhn segments, and $b$ is the Kuhn length. Within the framework of the solvent quality parameter $z$, good solvents are defined by $z>1$, theta solvents by $z \approx 0$, and poor solvents by $z<0$. For the purposes of this work, we are primarily interested in the good solvent regime such that $z$ $>$ 1.

\begin{figure}[t]
\centering
\includegraphics[height=8cm]{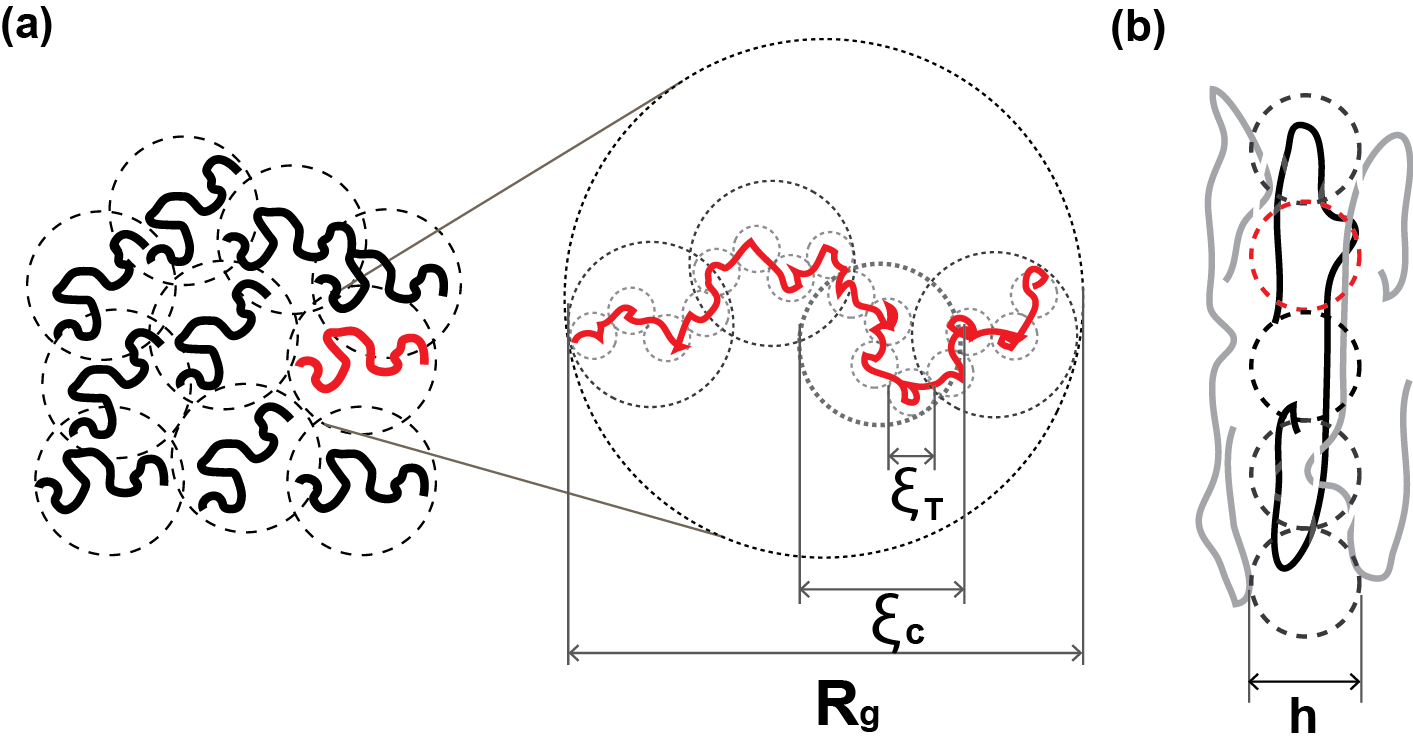}
  \caption{\label{fig:fig2}Semi-dilute polymer solutions in the context of the blob picture. (a) Schematic of a semi-dilute polymer solution near-equilibrium showing interpenetrating polymer coils and a single polymer chain in good solvent conditions. Characteristic length scales are the thermal blob size $\xi_T$, the concentration blob size $\xi_c$, and the radius of gyration $R_g$. (b) Schematic of a polymer chain under non-equilibrium conditions in an extensional flow in a semi-dilute solution. Under non-equilibrium flow conditions, a characteristic length scale $h$ can be defined as the average distance between neighboring chains.}  
\end{figure}

In semi-dilute polymer solutions, the near-equilibrium properties of polymer chains are determined by an interplay between {\em both} polymer concentration and solvent quality FIG.~\ref{fig:fig2} \cite{DeGennes1979, Rubinstein2003}. As polymer concentration is increased near the overlap concentration $c^*$, polymer chains begin to interpenetrate. Nevertheless, polymer volume fraction is relatively low for polymer concentrations near $c^*$, such that individual monomers are mainly surrounded by solvent. In order to make progress, blob theory can be used to describe the near-equilibrium properties of polymers for a given solvent quality and solution concentration in semi-dilute solutions \cite{Rubinstein2003,Jain2012}. As shown in FIG.~\ref{fig:fig2}a, the three characteristic length scales in semi-dilute polymer solutions are thermal blob size $\xi_T$, concentration blob size $\xi_c$, and radius of gyration $R_g$. The thermal blob size is defined as the length scale over which EV interactions effect chain size: 
\begin{equation} \xi_T = \frac{b^4}{\vert \text{v} \vert} = bN^{\frac{1}{2}}z^{-1} \end{equation}
On length scales smaller than $\xi_T$, EV interactions are weaker than thermal energy $k_BT$, and the conformations of thermal blobs are well described by an ideal random walk. In athermal solvents, $\text{v}=b^3$, and the thermal blob size is equal to the Kuhn step size $\xi_T = b$. The concentration blob size is defined as the length scale at which intermolecular interactions become relevant \cite{Jain2012}:
\begin{equation} \xi_c = bN^{\frac{1}{2}}\left( \frac{c}{c^*} \right)^{-\frac{\nu}{3\nu-1}}z^{2\nu-1} \end{equation}
where $\nu$ is the effective excluded volume exponent. On length scales larger than the thermal blob size $\xi_T$ but below the concentration blob size $\xi_c$, EV interactions are strong enough to swell the chain but are not yet screened by the surrounding chains, therefore, the conformations of concentration blobs are described by a self-avoiding walk. On length scales larger than the concentration blob size $\xi_c$, EV interactions are screened, and the conformation of the chain is a random walk of concentration blobs of size $\xi_c$. On these length scales, the end-to-end distance of a polymer is given as \cite{Jain2012}:
\begin{equation} \label{eq:endtoend} R = bN^{\frac{1}{2}} \left(\frac{c}{c^{\ast}} \right)^{-\frac{1}{2}\frac{2\nu-1}{3\nu-1}}z^{2\nu-1} \end{equation}
In the context of the blob model, the overlap concentration $c^{\ast}$ can be expressed as a function of solvent quality $z$ and polymer molecular weight (or number of Kuhn steps $N$) \cite{Jain2012}:
\begin{equation} c^{\ast} = b^{-3} N^{-\frac{1}{2}} z^{3-6\nu} \end{equation}
where $c^{\ast}$ is given in units of monomers per volume. Finally, as polymer concentration increases far above the overlap concentration $c \gg c^{\ast}$ and approaches the concentrated regime at $c^{\ast \ast}$, intramolecular EV interactions are gradually screened out, and the concentration blob size $\xi_c$ decreases until the size of concentration blob size is equal to the thermal blob size $\xi_c \approx \xi_T$. Here, as polymer concentration is increased above $c^{\ast \ast}$ into the concentrated regime, polymer chains are ideal on all length scales:
\begin{equation} c^{\ast \ast} = b^{-3} N^{-\frac{1}{2}} z \end{equation}
For concentrations $c > c^{\ast \ast}$, polymer chains are entangled at equilibrium. These expressions are used to plot the semi-dilute / concentrated boundary regime $c^{\ast \ast} / c^{\ast}$, as shown in FIG.~\ref{fig:fig1}.

In addition to the static conformational properties of polymer chains, the near-equilibrium dynamics of polymers in semi-dilute solutions can also be described using the blob model. The center-of-mass diffusion coefficient $D$ is given by the Einstein relation such that $D = k_BT / \zeta$, where $\zeta$ is the polymer friction coefficient. In the context of polymer chain dynamics, a hydrodynamic screening length $\xi_h$ can be defined as the length scale below which intramolecular hydrodynamic interactions (HI) are relevant. In terms of near-equilibrium properties, a reasonable assumption is to take the hydrodynamic screening to be equal to the concentration blob size $\xi_h \approx \xi_c$, which effectively means that EV and HI are screened at equivalent distances in semi-dilute solutions near equilibrium. The longest polymer relaxation time $\tau$ is given by the time scale required for a polymer coil to move a distance of its own size such that $\tau \approx R^2 / D \approx R^2\zeta / k_BT $. At length scales smaller than $\xi_h$ (or $\xi_c$), the relaxation time of a polymer segment of size $\xi$ follows the Zimm model: 
\begin{equation} \tau_{\xi} = \frac{\xi^3 \eta_s}{k_BT} = \frac{\eta_s b^3}{k_BT} N^{\frac{3}{2}} z^{6\nu-3} \left(\frac{c}{c^*} \right)^{-\frac{3\nu}{3\nu-1}} \end{equation} 
At length scales larger than $\xi_h$ (or $\xi_c$), intramolecular HI (and EV) is screened by surrounding polymer chains, and the chain size is a random walk of concentration blobs given by Eq. \ref{eq:endtoend}. The longest relaxation time of the polymer chain $\tau$ in semi-dilute solutions (good solvents) is: 
\begin{equation} 
\label{eq:relax}
\tau = \tau_{\xi} \left( \frac{N}{g}\right)^2 = \frac{\eta_s b^3}{k_BT} N^{3\nu} \left(\frac{c}{c^*}\right)^{\frac{2-3\nu}{3\nu-1}} (zN^{-\frac{1}{2}})^{6\nu-3} = \tau_{0} \left( \frac{c}{c^{\ast}}\right)^\frac{2-3\nu}{3\nu-1}z^{6\nu-3}
\end{equation} 
where $g$ is the number of steps in a concentration blob, $\eta_s$ is solvent viscosity, $\tau_0$ is the Rouse time or longest polymer relaxation time in dilute solution $\tau_{0} = \frac{\eta_s R_0^3}{k_BT}$, where $R_0 = N^{1/2}b$ is the polymer end-to-end distance in theta conditions. Note that the excluded volume exponent $\nu$ is a sensitive function of molecular weight and solvent quality $z$ (and therefore a sensitive function of $T$). Table \ref{tab:parameter} provides a summary of the scaling relations for polymer properties in semi-dilute solutions near equilibrium as arbitrary functions of solvent quality $z$ and concentration $c$.

\begin{table}[t]
 \begin{center}
	  \renewcommand{\arraystretch}{1.5}
   \begin{tabular}{| c | c | c |}
    \hline
        \rowcolor{gray!25}
    \textbf{Parameter} & \textbf{Dilute (good solvent)} & \textbf{Semi-dilute (good solvent)}   \\ \hline
    $\xi_T$ & $bN^{\frac{1}{2}}z^{-1}$ & $bN^{\frac{1}{2}}z^{-1}$   \\ \hline 
   $\xi_c$ &  --           & $bN^{\frac{1}{2}}\left( \frac{c}{c^*}\right)^{\frac{-\nu}{3\nu-1}}z^{2\nu-1}$   \\ \hline
    $R$       & $bN^{\frac{1}{2}}z^{2\nu -1}$ & $bN^{\frac{1}{2}} \left(\frac{c}{c^*}\right)^{-\frac{2\nu-1}{6\nu-2}}z^{2\nu-1}$ \\ \hline
    $\tau$  & $\frac{\eta_s b^3 }{k_BT}N^{\frac{3}{2}} z^{6\nu-3}$ & $\frac{\eta_s b^3 }{k_BT}N^{\frac{3}{2}} \left( \frac{c}{c^{\ast}}\right)^\frac{2-3\nu}{3\nu-1}z^{6\nu-3}$ \\ \hline
    \end{tabular}
    \caption{Relevant length and time scales for polymers in good solvent conditions in both dilute and semi-dilute regimes. Expressions are shown for arbitrary concentration $c$ and arbitrary solvent quality $z$.}
  	\label{tab:parameter}
  	\end{center}
\end{table}

Finally, the equilibrium blob picture changes drastically for polymer chains in non-equilibrium flows. Upon strong deformation, the relevant length scales and characteristic screening lengths for HI and EV are modified beyond their equilibrium scalings. In strong flow conditions, the concentration blob size depends on solvent quality, concentration, and flow strength. Moreover, an additional characteristic length scale should be considered and is related to the Pincus blob size \cite{Pincus1976}, denoted in FIG.~\ref{fig:fig2}b as ``h''. Recently, a new theoretical framework was developed to extend the blob model to non-linear flows of semi-dilute solutions \cite{Prabhakar2014}. The full theoretical description is complex and is beyond the scope of the present work.

\section{Experimental}
\subsection{Semi-dilute solution preparation}
We prepared a series of semi-dilute solutions of linear, double stranded DNA for single molecule studies. For all experiments, $\lambda$-phage DNA (Invitrogen, $48.5$ $\:$kbp, $M_w=3.2\times10^{7}$ Da, $\sim$0.5 mg/mL) is used, which is obtained as a buffered aqueous solution (10 mM Tris-HCl, pH 7.4, 0.1 mM EDTA, and 5 mM NaCl). Although stock $\lambda$-DNA solutions are provided at a nominally semi-dilute concentration ($\sim$12 $c^{\ast}$), we sought to increase the underlying solvent viscosity of the buffer solution in order to increase the longest polymer relaxation time. To this end, a method to gently mix concentrated $\lambda$-DNA solutions with a viscous sucrose buffer is developed, which results in a homogenous semi-dilute polymer solution in viscous buffer. 

We first measured the DNA concentration in the stock solutions using a UV-vis spectrophotometer (Nanodrop, Thermo Fisher). The DNA concentration in the stock solution was found to be in the range of 0.2-0.5 mg/mL, showing some variation between batches. Based on the measured stock solution concentration, a working volume of DNA solution that has a target corresponding mass of DNA to reach the target DNA concentration for semi-dilute solutions (Table \ref{tab:dnasamples}) is prepared. Next, the working volumes of stock DNA solutions are heated to 65$^{\circ}$C for 10 minutes, followed by snap cooling on ice to prevent concatemer formation. Stock DNA samples are slowly concentrated using a MiVac Quattro concentrator (Genevac) to a volume of 100 $\mu$L. Next, a viscous sucrose buffer (55$\%$ w/w sucrose, 30 mM Tris-HCl, 2 mM EDTA, 5 mM NaCl, pH 8.0) is added to the concentrated DNA sample to yield a solution with final working volume of 1.0 mL. This procedure allows us to prepare semi-dilute DNA solutions while controlling the volume of aqueous buffer in the working DNA solutions, thereby enabling control over the final solvent viscosity $\eta_s$ for microfluidics experiments. Solution viscosities are measured using a benchtop viscometer (Brookfield) at 22$^{\circ}$C. In general, we aimed to achieve a target solvent viscosity of $\sim$50 cP, though the longest polymer relaxation time was measured using direct single molecule imaging for each solution separately (Section IV). 

In order to ensure sample homogeneity, semi-dilute DNA solutions were subjected to a series of repeated heat and mix cycles prior to single molecule experiments. Here, samples were gently heated to 55$^{\circ}$C for 10-15 minutes, followed by rotational mixing of sample vials at room temperature for 10 minutes. This procedure is repeated for 10 cycles, followed by rotational mixing overnight at 4$^{\circ}$C. Following solution preparation, DNA concentration is measured using a UV-vis spectrophotometer (Nanodrop, Thermo Fisher). DNA solution concentrations are determined by measuring absorbance at a wavelength of 260 nm and using an extinction coefficient of $\epsilon$ = 0.020 mL $\mu$g$^{-1}$ cm$^{-1}$. Agarose gel electrophoresis was also used to assess the quality and integrity of DNA samples from semi-dilute solutions post-mixing in order to ensure that sample degradation does not occur prior to experimentation. In all cases, gels showed a clear band at the expected molecular weight relative to a control sample of stock $\lambda$-DNA, with no fragments shorter or longer than $\lambda$-DNA.

Using this method, a series of semi-dilute DNA solutions with concentrations spanning above and below $c^{\ast}$ (Table \ref{tab:dnasamples}) were prepared. We used an overlap concentration $c^{\ast} \approx$ 40 $\mu$g/mL for $\lambda$-DNA based on the previously reported value of $R_g \approx$ 0.6 $\mu$m for unlabeled $\lambda$-DNA in aqueous buffer, which was determined using a combination of dynamic light scattering to determine the hydrodynamic radius $R_H$ and a rigorous parameter matching scheme based on Brownian dynamics simulations \cite{Pan2014}. All experiments are conducted with a circulating water bath to maintain a constant temperature in the microdevice at $T$ = 22 $^{\circ}$C, which is above the theta temperature of $T_{\theta}$ = 14$^{\circ}$C determined by static light scattering \cite{Pan2014}. Based on these conditions, all experiments are performed in the good solvent regime for double stranded DNA in aqueous solution \cite{Pan2014}. Using this approach, several solutions with $\lambda$-phage DNA concentrations ranging from ultra-dilute to semi-dilute (Table~\ref{tab:dnasamples}) were prepared: 10$^{-5}$ $c^{\ast}$ (0.4 ng/mL), 0.5 $c^{\ast}$ (20 $\mu$g/mL), 1.25 $c^{\ast}$ (50 $\mu$g/mL), $c^{\ast}$ (56 $\mu$g/mL), 2 $c^{\ast}$ (80 $\mu$g/mL), 2.7 $c^{\ast}$ (108 $\mu$g/mL), 3.6 $c^{\ast}$ (144 $\mu$g/mL).

\begin{table}[t]
	\centering
    \begin{tabular}{| c | c | c | c | c | c | c | c |}
    \rowcolor{gray!25}
    \hline
    $\lambda$-DNA concentration ($c^*$) & $10^{-5}$ $c^*$ & 0.5 $c^*$ & 1.25 $c^*$ & 1.4 $c^*$ & 2.0 $c^*$ & 2.7 $c^*$ & 3.6 $c^*$ \\ 
    \hline
    $\lambda$-DNA concentration ($\mu$g/mL) & $10^{-4}$  & 20  & 50 & 56 & 80  & 108  & 144 \\ 
    \hline   
    \end{tabular}
    \caption{Semi-dilute DNA solutions used for this work.}
  	\label{tab:dnasamples}
\end{table}

For single molecule imaging, a small amount of fluorescently labeled $\lambda$-DNA is added to an unlabeled background solution of semi-dilute DNA. To prepare fluorescently labeled $\lambda$-DNA, stock YOYO-1 solution (10$^{-3}$ M, Molecular Probes) is diluted to a concentration of 10$^{-5}$ M YOYO-1 in imaging buffer (30 mM Tris-HCl, 2 mM EDTA, 5 mM NaCl, pH 8). Separately, stock $\lambda$-phage DNA was diluted to 10 $\mu$g/mL in imaging buffer and subsequently heated to 65$^{\circ}$C for 10 minutes, followed by snap cooling to prevent concatemer formation. Next, the diluted $\lambda$-DNA solution was mixed with the diluted YOYO-1 solution in imaging buffer to achieve a final DNA concentration of 1 $\mu$g/mL in the staining solution. Using this approach, DNA was labeled at a ratio of 1 dye per 4 base pairs. The DNA/dye solution was incubated for 1.5 hours at room temperature in the dark before use. 

Following DNA staining, fluorescently labeled DNA was added to the unlabeled semi-dilute DNA solution to achieve a final concentration of $\sim$10$^{-4}$ $\mu$g/mL labeled `probe' DNA in the semi-dilute solution background. In addition, small amounts of the reducing agent $\beta$-mercaptoethanol (6 $\mu$L/mL) and an oxygen scavenging enzyme system based on glucose oxidase (1.5 $\mu$L/mL), catalase (1.5 $\mu$L/mL), and (6 $\mu$L/mL) $\beta\text{-}D\text{-}$glucose (1$\%$ w/w) was added to enhance photostability. The volume change after addition of these reagents is 1.5$\%$, yielding a negligible change in polymer concentration. Finally, the semi-dilute solution containing fluorescently labeled DNA and photobleaching reagents was rotationally mixed for 40 minutes at room temperature prior to imaging. 

\subsection{Optics, Imaging, and Microfluidic Devices} 

\begin{figure}
\centering
\includegraphics[height=15cm]{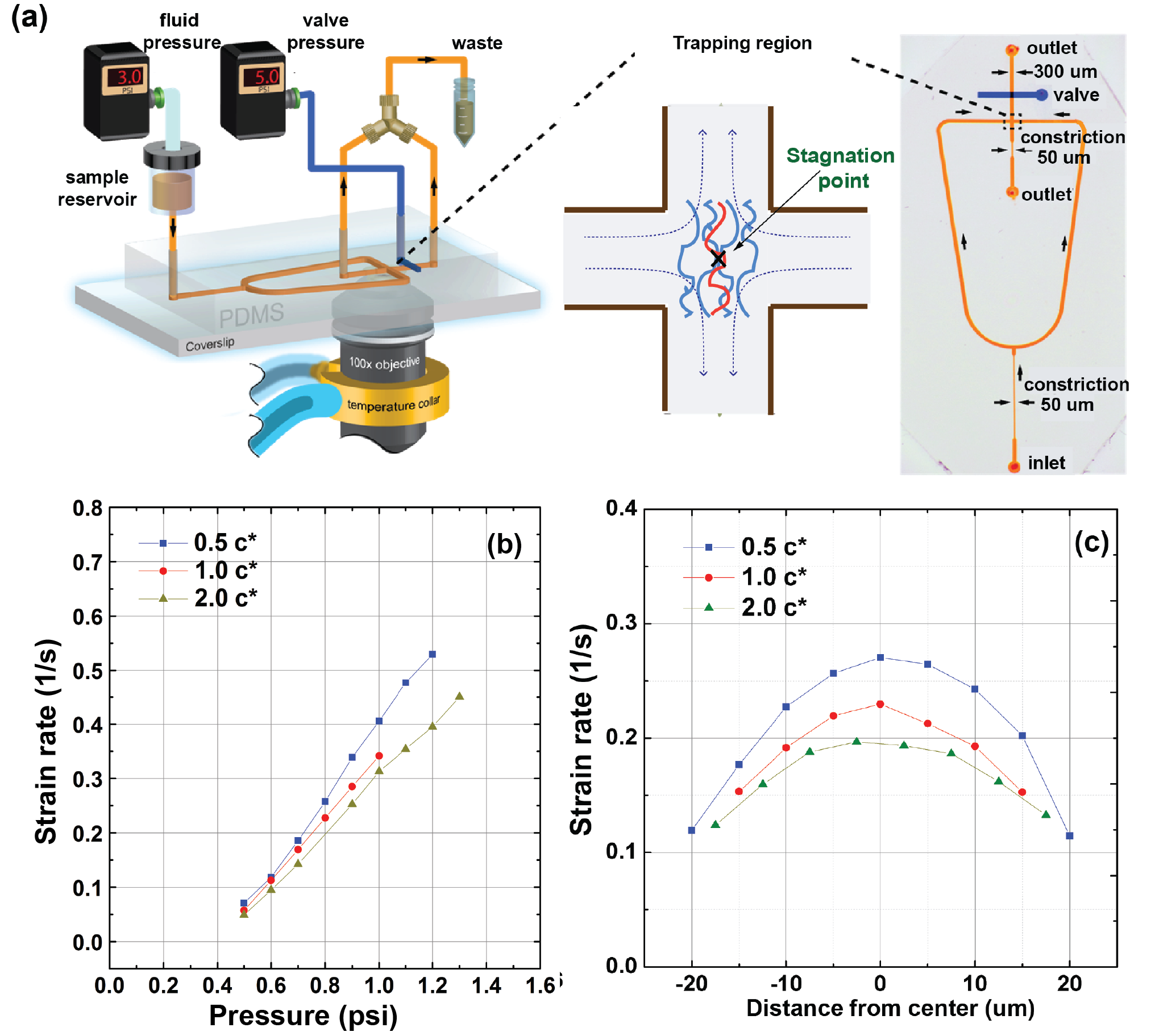}
  \caption{\label{fig:fig3}Cross-slot microfluidic device and strain rate calibration. (a)$\:$ Schematic of the cross-slot microfluidic device used to generate planar extensional flow for single molecule imaging. (b) Strain rate calibration in a cross-slot device at the mid-plane as a function of inlet pressure. Bead tracking experiments are performed in three different semi-dilute solutions with polymer concentration 0.5 $c^{\ast}$, 1 $c^{\ast}$, and 2 $c^{\ast}$. (c) Strain rate calibration as a function of distance from the horizontal mid-plane in the device.}   
\end{figure}

Single polymer dynamics were observed in a planar extensional flow generated in a PDMS-based microfluidic device with a cross-slot channel design. Here, two opposing laminar streams converge at the cross-slot junction and exit through mutually perpendicular outlet channels, thereby creating a planar extensional flow, which is a two-dimensional flow containing a fluid stagnation point (zero-velocity point). A custom hydrodynamic trap was used to enable the direct observation of chain dynamics in planar extensional flow with a defined strain rate $\dot{\epsilon}$ for a finite observation time. The hydrodynamic trap is based on active feedback control of a stagnation point flow generated at the cross-slot junction in a PDMS-based microfluidic device FIG.~\ref{fig:fig3}. The full details of the hydrodynamic trap have been previously reported in prior work \cite{Tanyeri2011}; in brief, an on-chip membrane valve is used to modulate the fluidic resistance in one outlet channel, thereby enabling control over the stagnation point position and effective trapping of single polymers for long times. The action of the valve enables the trapping of single polymers, and under the flow rates used in this study, the valve action results in negligible changes in the strain rate $\dot{\epsilon}$ during an experiment.

Cross-slot microfluidic devices were fabricated using standard methods in multi-layer soft lithography. In brief, a two-layer PDMS device is fabricated containing a fluidic layer positioned below a control (valve) layer. An optical micrograph of a sample device is shown in FIG.~\ref{fig:fig3}a. Two separate master molds (one each for the fluidic and control layers) were first fabricated using SU-8 photoresist (Microchem) patterned onto silicon wafers. PDMS was mixed in 15:1 and 5:1 base to cross-linker ratios for the fluidic and control layer, respectively. The two layers were partially cured at 65$^{\circ}$C for 25-30 minutes, and control layer was later aligned with the fluidic layer. Next, the two layers were cured for an additional 2-4 hours. After the final curing step, the remaining fluid inlet and outlet holes were punched, and the PDMS devices were bonded to glass coverslips after oxygen plasma cleaning.

A schematic depicting the experimental setup is shown in FIG.~\ref{fig:fig3}a. In this work, pressure driven flow is used to generate fluid flow in cross-slot microdevices. In particular, we designed microdevices with extended inlet channels and a constriction region in the inlet channel (with 50 $\mu$m channel width), which effectively allowed for working fluid pressures between 1-3 psi for 20-100 cP solution viscosities. An on-chip membrane valve was positioned above one of the outlet channels and equal distance from a constriction region in the opposite outlet channel relative to the cross-slot. Applying pressure to this valve constricts the outlet channel in the fluidic layer underneath the control layer, which can be used to effectively manipulate the stagnation point position using feedback control. The viewing solution containing fluorescently labeled DNA was introduced into the PDMS device via a sample tube connected to a pressure transducer (Proportion Air). Pressure driven flow was used to enable precise control over flow rate, thereby allowing for rapid start-up and shutdown of the flow. Characteristic time scales for step changes in flow rate were found to be $<$ 1 sec, which is significantly less than the duration of transient polymer stretching events. A custom cooling jacket was fitted to the microscope objective, thereby enabling precise temperature control of the viewing solution via circulating water bath.

Single molecule imaging and detection was performed using an inverted epifluorescence microscope (IX-71, Olympus) coupled to an electron multiplying charge coupled device (EMCCD) camera (Andor iXon). Fluorescently labeled DNA samples were illuminated by a solid-state CW laser laser (Coherent, 488 nm) and imaged using a 1.45 NA, 100$\times$ oil immersion objective lens. Images were acquired using an additional 1.6$\times$ magnification lens in the optical path prior to the EMCCD camera. Full frame images (512 $\times$ 512 pixels) were acquired at a frame rate of 30 Hz. 

\section{Results and Discussion}
\subsection{Flow field characterization in semi-dilute solutions}

We first characterized flow field kinematics in microfluidic cross-slot devices using particle tracking (FIG.~\ref{fig:fig3}b,c). Experimental characterization of flow fields in semi-dilute polymer solutions is essential to ensure that flow fields are well behaved and that polymer samples are homogenous in composition. For these experiments, 0.84 $\mu$m diameter fluorescent beads (SpheroTech) are introduced into a series of semi-dilute polymer solutions, and we performed particle tracking experiments using three different concentrations of polymer: 0.5 $c^{\ast}$, 1 $c^{\ast}$, and 2 $c^{\ast}$. Solutions were viscosity matched to those used in DNA trapping experiments for accurate determination of fluid strain rates. Images were captured using a CCD camera (AVT Stingray) at frame rates of at least 60 Hz. Individual particle trajectories were tracked and mapped using the ParticleTracker plugin for ImageJ. From particle position data, instantaneous bead velocities were determined, and data were fit using a non-linear least squares algorithm to following relationship for planar extensional flow: 
\[\begin{bmatrix}v_x \\ v_y \end{bmatrix} = \begin{bmatrix} \dot{\epsilon} & 0 \\ 0 & -\dot{\epsilon} \end{bmatrix} \begin{bmatrix} x - x_0 \\  y - y_0 \end{bmatrix}\] where $v_x$, $v_y$, $x$, and $y$ are velocities and positions in the $x$ and $y$ directions, respectively (known quantities), and $\dot{\epsilon}$, $x_0$, and $y_0$ are fitting parameters. Here, $\dot{\epsilon}$ is the fluid strain rate and ($x_0$, $y_0$) is the stagnation point position (unknown quantities).

We first determined the strain rate near the center of the cross-slot device as a function of pressure (via pressure-driven flow). Strain rate increases linearly with pressure over the characteristic range of strain rates used for single polymer dynamics (FIG.~\ref{fig:fig3}b). Upon increasing polymer concentration, the strain rate slightly decreases, which is suggestive of flow field modification away from a simple Newtonian solvent, similar to prior work on flow birefringence of synthetic polymers in semi-dilute solutions in extensional flow \cite{Dunlap1987}. In addition, we also determined the flow profile as a function of distance away from the horizontal mid-plane in the $z$-direction, which is the stagnant (no flow) direction (FIG.~\ref{fig:fig3}c). Here, a near parabolic flow profile is observed with pronounced flattening upon increasing polymer concentration. Bulk rheological measurements on semi-dilute unentangled DNA solutions ranging in concentration from 1 $c^{\ast}$ to 10 $c^{\ast}$ have been extensively carried out by Prakash and coworkers \cite{Panthesis2014}, including both linear viscoelastic measurements and steady shear rheology. For semi-dilute solutions of lambda DNA at 1.5 $c^{\ast}$ and 2.3 $c^{\ast}$, the onset of shear thinning was observed to occur around $Wi$=1.0, so it is possible that the flattening of the velocity profile observed in the 2 $c^{\ast}$ polymer solution in the cross-slot device could arise due to mild shear thinning at $Wi \approx$ 2-3, which corresponds to the maximum $Wi$ based on these strain rates.

\subsection{Longest polymer relaxation time}

\begin{figure}[t]
\centering
\includegraphics[height=8cm]{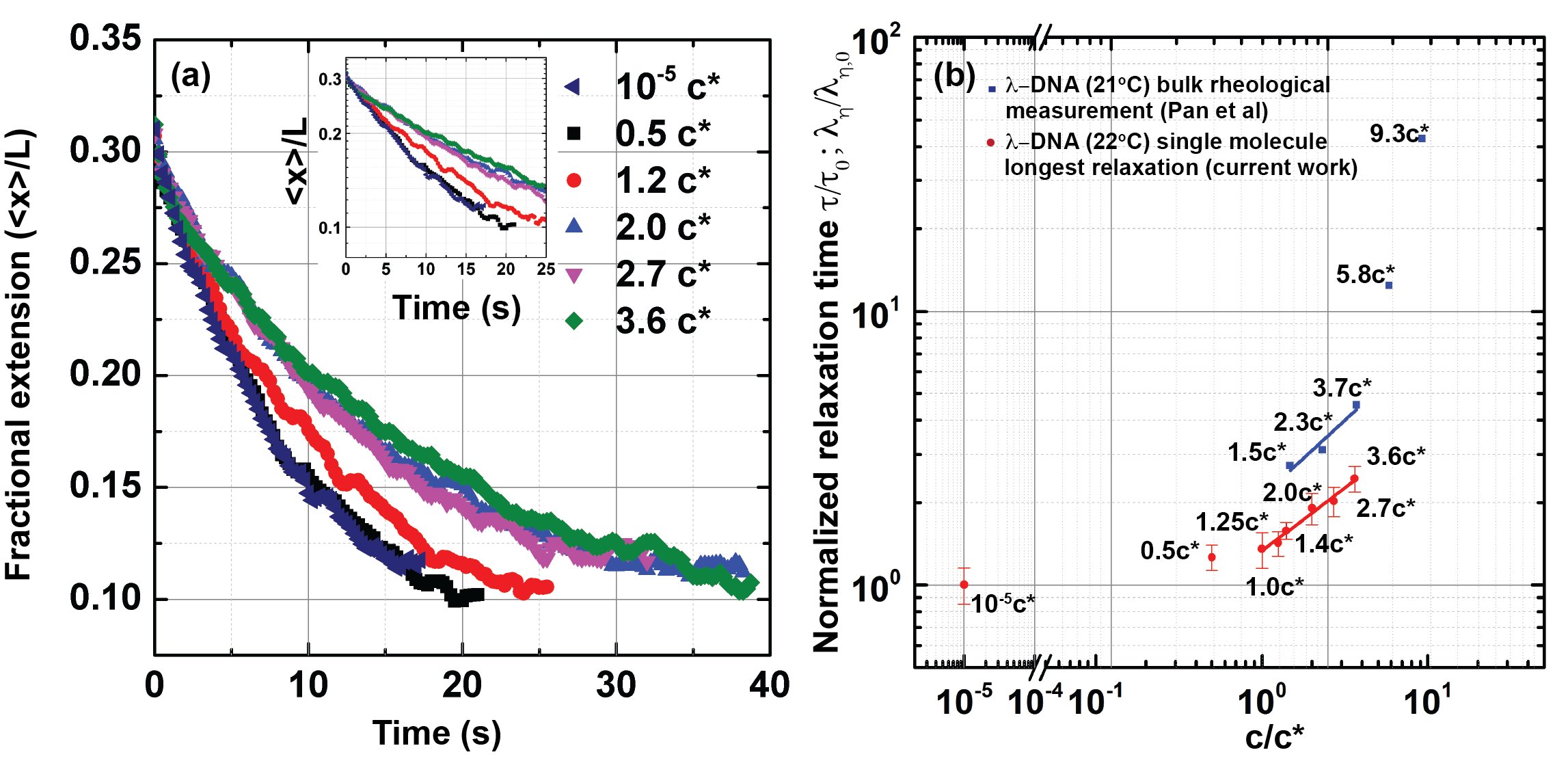}
  \caption{\label{fig:bulk}Longest relaxation time of DNA in semi-dilute solutions. (a) Ensemble average of single molecule relaxation trajectories at several different concentrations ($N \approx$ 40 molecules in each ensemble). Inset: semi-log plot of polymer relaxation trajectories. (b) Normalized longest relaxation times as a function of scaled concentration $c / c^{\ast}$. Longest polymer relaxation times $\tau$ are normalized to the corresponding dilute solution relaxation times $\tau_0$ at the same solvent viscosity. (circles) Normalized longest relaxation times from single molecule experiments on semi-dilute $\lambda$ DNA solutions as a function of scaled polymer concentration. Error bars correspond to the standard deviation of longest relaxation times from the molecular ensemble at each concentration. (squares) Normalized relaxation times from bulk rheological data on semi-dilute $\lambda$ DNA solutions, where zero-shear viscosity measurements are used to determine a longest polymer relaxation time \cite{Pan2014}.}
\end{figure}

Following flow field characterization, we embarked on single polymer dynamics experiments. We first studied the longest conformational relaxation time of polymers in semi-dilute solutions following cessation of extensional flow (FIG.~\ref{fig:bulk}). In this experiment, a semi-dilute polymer solution doped with fluorescently labeled $\lambda$-DNA is flowed into the cross-slot device at a fairly high flow rate $Wi>1$, followed by abrupt stoppage of fluid flow. We then observe the relaxation process of single stretched polymers from high extension. Image analysis software is used to track the transient extension $x$ of single polymers following cessation of flow, as shown in FIG.~\ref{fig:bulk}a. In particular, we track the maximum polymer extension $x$ along the principal axis of extension, which can be considered as the maximum polymer extension projected onto the extensional axis. Relaxation times are determined by fitting the terminal 30$\%$ of projected fractional extension $x/L$ to a single exponential decay: $\langle x \cdot x \rangle /L^2 = A \: \exp(-t / \tau) + B$, where $\tau$ is the longest relaxation time and $A$ and $B$ are the fitting constants. A semi-log plot of polymer relaxation is shown in the inset of FIG.~\ref{fig:bulk}a, where a clear linear relation is observed for the terminal 30\% of polymer relaxation. We further compared our single molecule DNA relaxation data in semi-dilute solutions (based on single chain conformational relaxation time $\tau$) to bulk experimental data on relaxation of semi-dilute DNA solutions (based on zero-shear viscosity data used to determine a relaxation time $\lambda_{\eta}$). Results are shown in FIG.~\ref{fig:bulk}b, which plots the normalized longest relaxation times $\tau / \tau_0$ and $\lambda_{\eta} / \lambda_{\eta,0}$ as functions of the normalized concentration $c / c^{\ast}$ \cite{Pan2014}. Here, the longest relaxation time ($\tau$ or $\lambda_{\eta}$) for each semi-dilute solution is normalized to the longest relaxation time of the corresponding dilute solution ($\tau_0$ or $\lambda_{\eta,0}$) at an equivalent solvent viscosity $\eta_s$, where $\tau_0$ is obtained by single molecule experiments in the ultra-dilute limit.

The normalized single molecule and bulk relaxation data in FIG.~\ref{fig:bulk}b are both consistent with a power law scaling as a function of scaled concentration $c / c^{\ast}$. Based on Eq. \ref{eq:relax}, we expect that the longest polymer relaxation time in semi-dilute unentangled solutions follows the power law scaling $\tau / \tau_0 \sim \left( c / c^{\ast} \right)^\frac{2-3\nu}{3\nu-1}$, where $\nu$ is the effective excluded volume exponent. We found that our single molecule data were consistent with a power law scaling $\tau / \tau_0 \sim ( c / c^* )^{0.48}$, which yields $\nu =$ 0.56. In fact, we found that $\nu$ lies between 0.53 and 0.56 given the uncertainty in the experimental relaxation times determined in this work. We also compared our single molecule data with bulk experimental data by Pan et al. \cite{Pan2014}, who measured the zero-shear viscosity of $\lambda$-DNA at $T$ = 21$^{\circ}$C, which can be used to determine a longest relaxation time $\lambda_{\eta}$ using the relation $\lambda_{\eta} = M \eta_{p0} / cN_Ak_BT$, where $\eta_{p0}$ is the zero-shear viscosity. In order to compare to single molecule experiments, we plot the bulk relaxation data normalized to the longest relaxation time in dilute solutions such that $\lambda_{\eta}/\lambda_{\eta,0} = \eta_{p0}/c[\eta]_0\eta_s$, where $[\eta]_0$ is the zero-shear intrinsic viscosity \cite{HCttinger1996}. For this comparison, we take $[\eta]_0$ = 11.9 mL/mg for $\lambda$-DNA in the range of 21-25$^{\circ}$C \cite{Tsortos2011}. Using this approach, we find that bulk experimental data on $\lambda$-DNA relaxation is consistent with the power law scaling determined in our single molecule relaxation data in the same concentration regime for semi-dilute unentangled polymer solutions, where single molecule measurements show $\tau / \tau_0 \sim ( c / c^* )^{0.48}$ (thereby giving $\nu$ = 0.56) and bulk rheology measurements yield $\lambda_{\eta} / \lambda_{\eta,0} \sim ( c / c^* )^{0.54}$ (thereby giving $\nu$ = 0.55). Previous studies on single molecule relaxation of T4 DNA (165.6 kbp) also show a similar scaling exponent $\tau/\tau_0 \sim ( c / c^* )^{0.5}$ (which gives $\nu$ = 0.56) \cite{Liu2009}. These results are all fairly consistent for DNA.

Despite the good agreement in scaling between bulk and single molecule relaxation data, the effective excluded volume exponent $\nu =$ 0.56 is lower than expected for flexible polymers in the good solvent regime ($\nu \approx$ 0.588) \cite{Rubinstein2003}. We can rationalize this result using several physical arguments. First, our experiments are performed at temperature $T$ = 22$^{\circ}$C, which is larger than the theta temperature for DNA in aqueous solutions ($T_{\theta}$ = 14$^{\circ}$C). Nevertheless, our experiments correspond to the cross-over region in solvent quality between theta and very good solvents, and we therefore expect that the excluded volume exponent will be less than 0.588. Secondly, Prakash and coworkers \cite{Pan2014a} recently showed that the behavior of synthetic worm-like chains and DNA can be well described by taking semi-flexibility into account in the definition of the solvent quality parameter. Using this approach, the apparent semi-flexibility of a polymer can be directly accounted for in solvent quality. Additional work further supports the notion that polymer flexibility may impact the effective excluded volume exponent $\nu$. Recently, Tree et al. \cite{Tree2013} used chain-growth Monte Carlo simulations based on a pruned-enriched Rosenbluth method (PERM) to study the effect of local polymer flexibility on the global properties of polymer chains in athermal solvents. In particular, these authors examined the effect of monomer aspect ratio $b / d$ on equilibrium chain dimensions, where $b$ is the Kuhn length and $d$ is the effective chain width. Results from PERM simulations show that the effective excluded volume exponent for $\lambda$-DNA in an athermal solvent is $\nu \approx$ 0.55, which is less than expected for the theoretical value for flexible chains in an athermal solvent due to local chain flexibility \cite{Tree2013}. Moreover, Krichevsky and coworkers recently studied DNA chain conformation using scanning fluorescence correlation spectroscopy, revealing an excluded volume exponent $\nu \approx $ 0.52 for DNA in aqueous solution \cite{Nepal2013}. Based on these results, we might not expect to observe excluded volume exponents similar to truly flexible polymers regardless of solvent quality in the good solvent regime, largely due to the semi-flexible nature of double stranded DNA. Taken together, we conclude that the excluded volume exponent for $\lambda$-DNA appears to be in the cross-over regime between theta solvents and athermal solvents, though this is likely a reflection of both polymer flexibility and solvent quality.

\begin{figure}
\centering
\includegraphics[height=16cm]{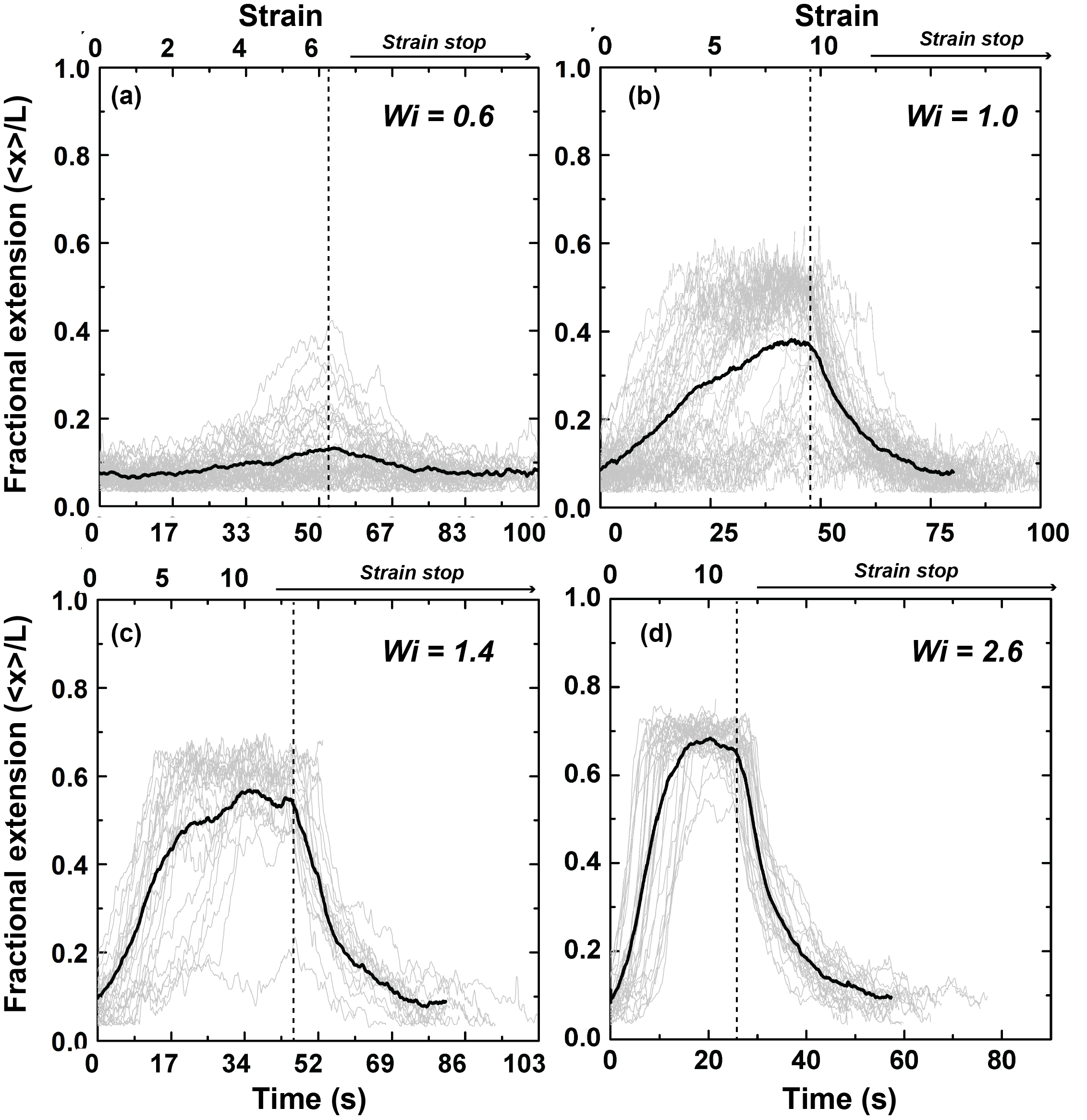}
  \caption{\label{fig:fig5}Transient polymer stretch in a step strain rate experiment in planar extensional flow. Results are shown for dynamics in 1 $c^{\ast}$ solutions as a function of increasing flow strength: (a) $Wi$ = 0.6, (b) $Wi$ = 1.0, (c) $Wi$ = 1.4, (d) $Wi$ = 2.6. Thin traces show individual molecular stretching trajectories, and thick traces show ensemble average stretch. The dotted lines indicate where the step-strain rate is stopped.}  
\end{figure}

\subsection{Transient and steady-state dynamics in extensional flow} 

\begin{figure}
\centering
\includegraphics[height=16cm]{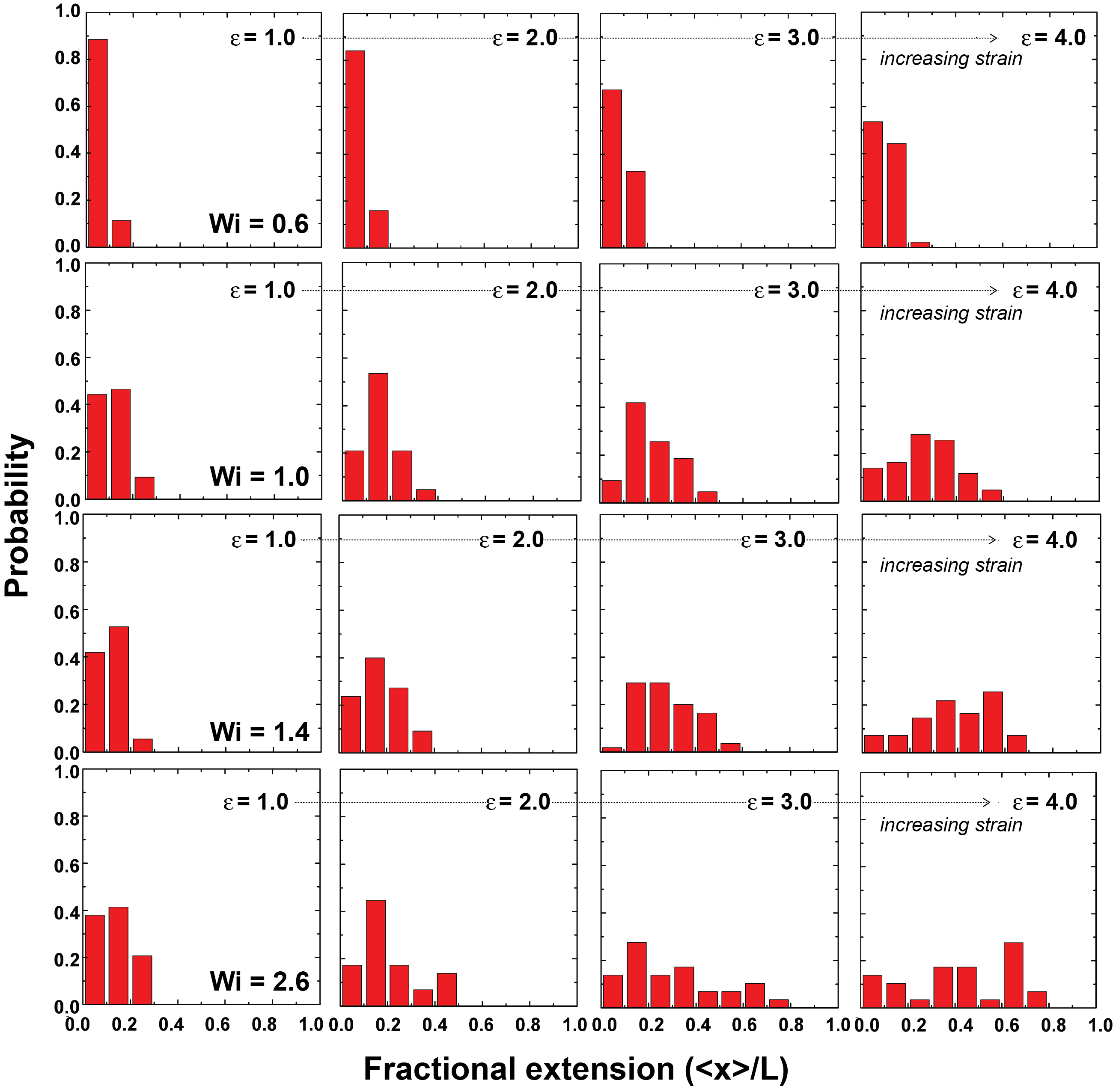}
  \caption{\label{fig:fig6}Probability distribution of chain extension in semi-dilute ($1\:c^{\ast}$) solutions flows in planar extensional flow. Distributions are shown for a total accumulated strain of $\epsilon$ = 1.0, 2.0, 3.0, and 4.0 across several different flow strengths $Wi$ = 0.6, 1.0, 1.4, and 2.6.}   
\end{figure} 

We next studied the non-equilibrium stretching dynamics of single polymers in semi-dilute solutions in planar extensional flow. In these experiments, a step input on the strain rate $\dot{\epsilon}$ is applied, and polymer solutions are subjected to an extensional flow characterized by a Weissenberg number $Wi = \tau \dot{\epsilon}$ for a finite amount of accumulated fluid strain $\epsilon = \int_{0}^{t_{obs}}\dot{\epsilon}\: dt$, which is known as the Hencky strain (FIG.~\ref{fig:fig5}). Here, $t_{obs}$ is defined to be the duration of step-strain rate deformation on the polymer sample. Using the feedback-controlled hydrodynamic trap, we are able to probe polymer dynamics in precisely controlled extensional flows with constant $Wi$. In this way, we explore the non-linear, transient dynamics of semi-dilute solutions during a step strain rate input, which includes transient dynamics during start up and following the cessation of flow. In these experiments, single fluorescently labeled polymers are first allowed to relax for several relaxation times $\tau$ under no flow conditions. Next, a step strain rate at time $t$ = 0 is imposed, and single polymers are imaged at a precisely controlled $Wi$ for a finite amount of strain $\epsilon$. Finally, the flow is halted, and the polymer relaxes to back to an equilibrium coiled state.

\begin{figure}
\centering
\includegraphics[height=6cm]{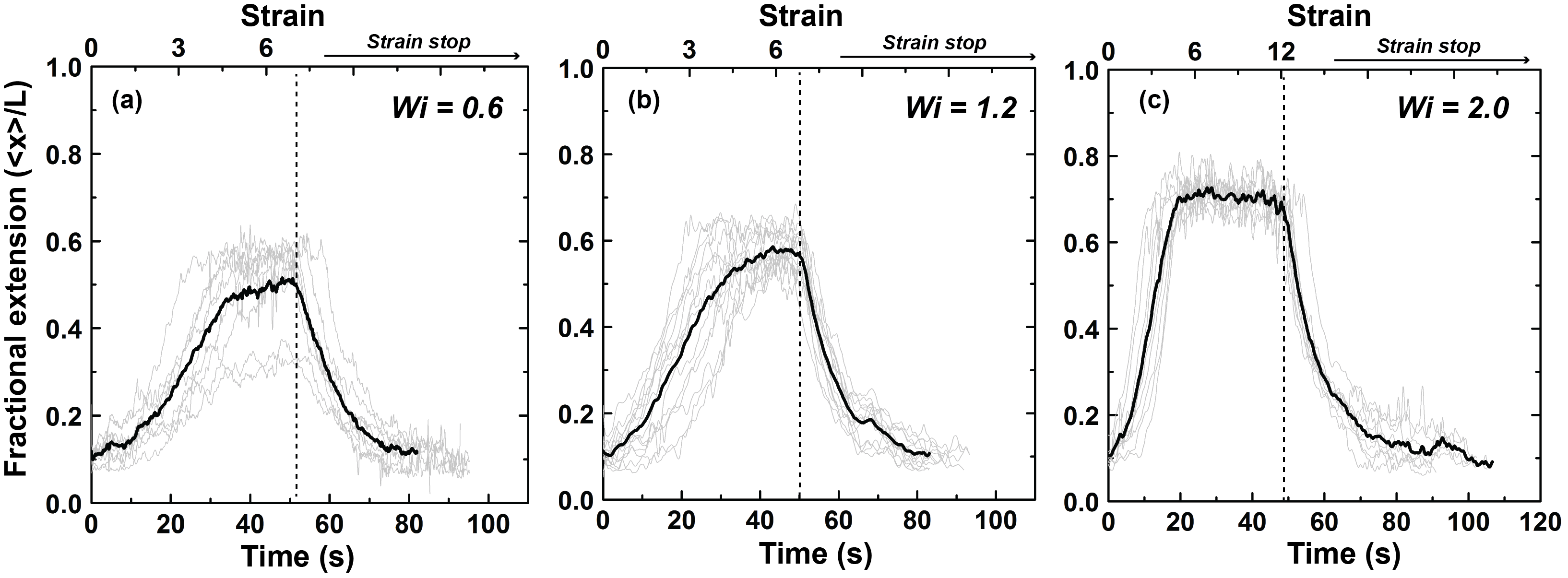}
  \caption{\label{fig:fig7}Transient polymer stretch in a step strain rate experiment in planar extensional flow. Results are shown for dynamics in ultra-dilute ($10^{-5}c^*$) polymer solutions as a function of increasing flow strength: (a) $Wi$ = 0.6, (b) $Wi$ = 1.2, (c) $Wi$ = 2.0. Thin traces show individual molecular stretching trajectories, and thick traces show ensemble average stretch. The dotted lines indicate where the step-strain rate is stopped.}  
\end{figure}

\begin{figure}
\centering
\includegraphics[height=13cm]{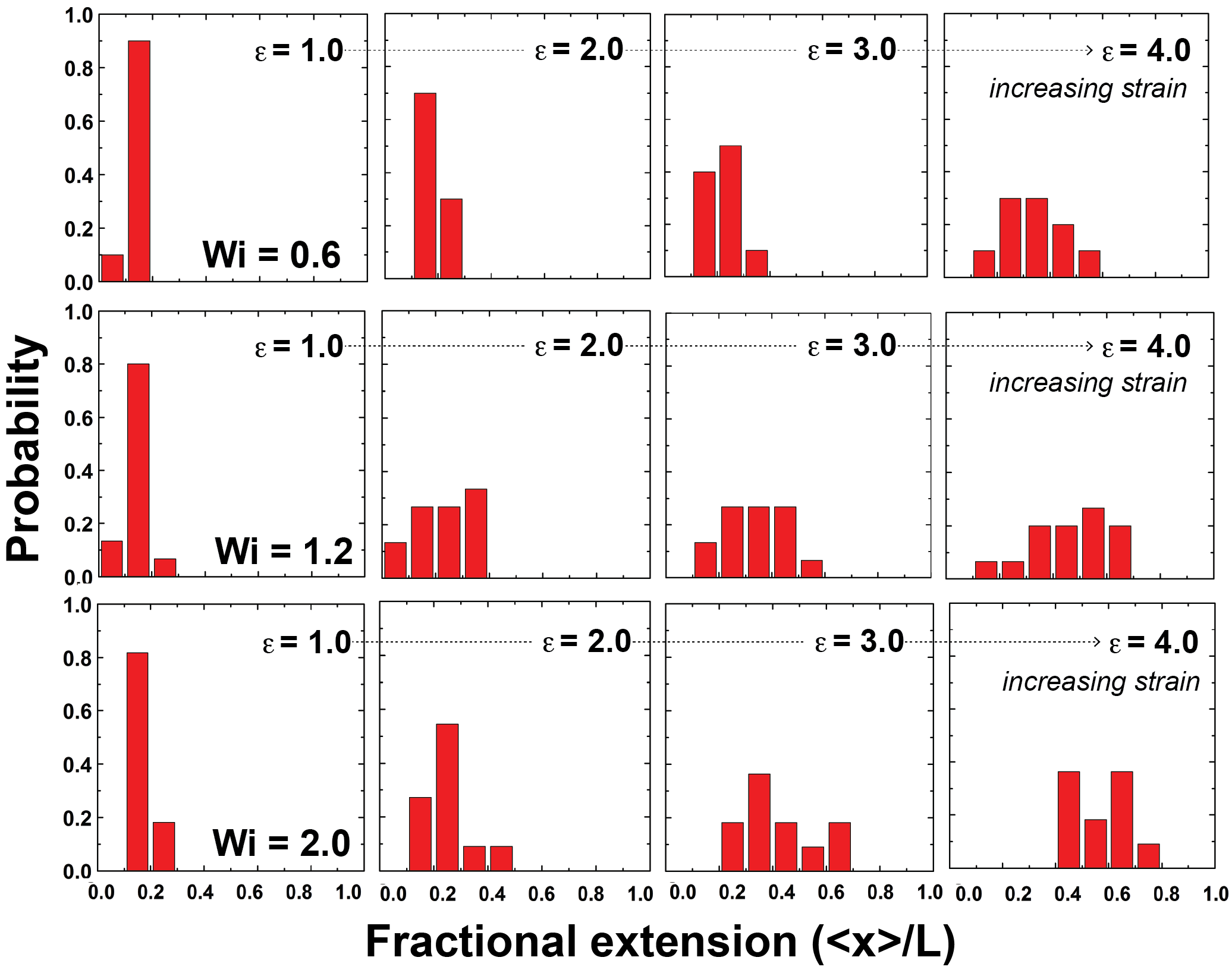}
  \caption{\label{fig:fig8}Probability distribution of chain extension in ultra-dilute ($10^{-5}c^*$) solutions flows in planar extensional flow. Distributions are shown for a total accumulated strain of $\epsilon$ = 1.0, 2.0, 3.0, and 4.0 across several different flow strengths $Wi$ = 0.6, 1.2, and 2.0.}   
\end{figure}

Using this approach, we studied the dynamics of single DNA molecules in semi-dilute solutions (1 $c^{\ast}$) at $Wi$ = 0.6, 1.0, 1.4 and 2.6, where $Wi$ is defined using the longest polymer relaxation time in 1 $c^{\ast}$ solutions. Transient fractional extension for semi-dilute solutions is shown in FIG.~\ref{fig:fig5}, and the corresponding probability distributions of polymer extension are shown in FIG.~\ref{fig:fig6}. The transient stretching data in FIG.~\ref{fig:fig5} show both the individual single molecule stretching trajectories and the ensemble average for each $Wi$. Across all $Wi$, the minimum accumulated fluid strain was $\epsilon$ = 6, and ensemble averages are determined from a minimum of 30-50 individual trajectories. For comparison, we also performed a series of experiments to study transient polymer stretching in ultra-dilute solutions (10$^{-5}$ $c^{\ast}$) under similar flow strengths for a step strain-rate input in planar extensional flow. Transient fractional extension for ultra-dilute solutions is shown in FIG.~\ref{fig:fig7}, and the corresponding probability distributions of polymer extension in dilute solutions are shown in FIG.~\ref{fig:fig8}. In this way, it is possible to directly compare transient dynamics in ultra-dilute and semi-dilute solutions in planar extensional flow.

Strikingly, a broad variability in transient stretching dynamics is observed within the distribution of trajectories for semi-dilute solutions (FIG.~\ref{fig:fig6}). Here, we observe a broad distribution in the {\em onset} of stretching in transient extensional flow in semi-dilute polymer solutions. The distribution broadens as the accumulated strain increases, indicating the presence of molecular individualism in start-up of extensional flow. In comparing these results to dilute solution dynamics, semi-dilute solutions clearly show a much broader distribution of transient polymer extension (FIG.~\ref{fig:fig6}) compared to ultra-dilute solutions under similar flow strengths and accumulated fluid strains (FIG.~\ref{fig:fig8}). A broad probability distribution of polymer extension is not well described by a Gaussian function, which is the characteristic configurational distribution function for polymer stretch in dilute solution extensional flows from kinetic theory \cite{Bird1977}. We conjecture that the broad distribution in polymer extension arises in semi-dilute solutions due to intermolecular interactions.

\begin{figure}
\centering
\includegraphics[height=15cm]{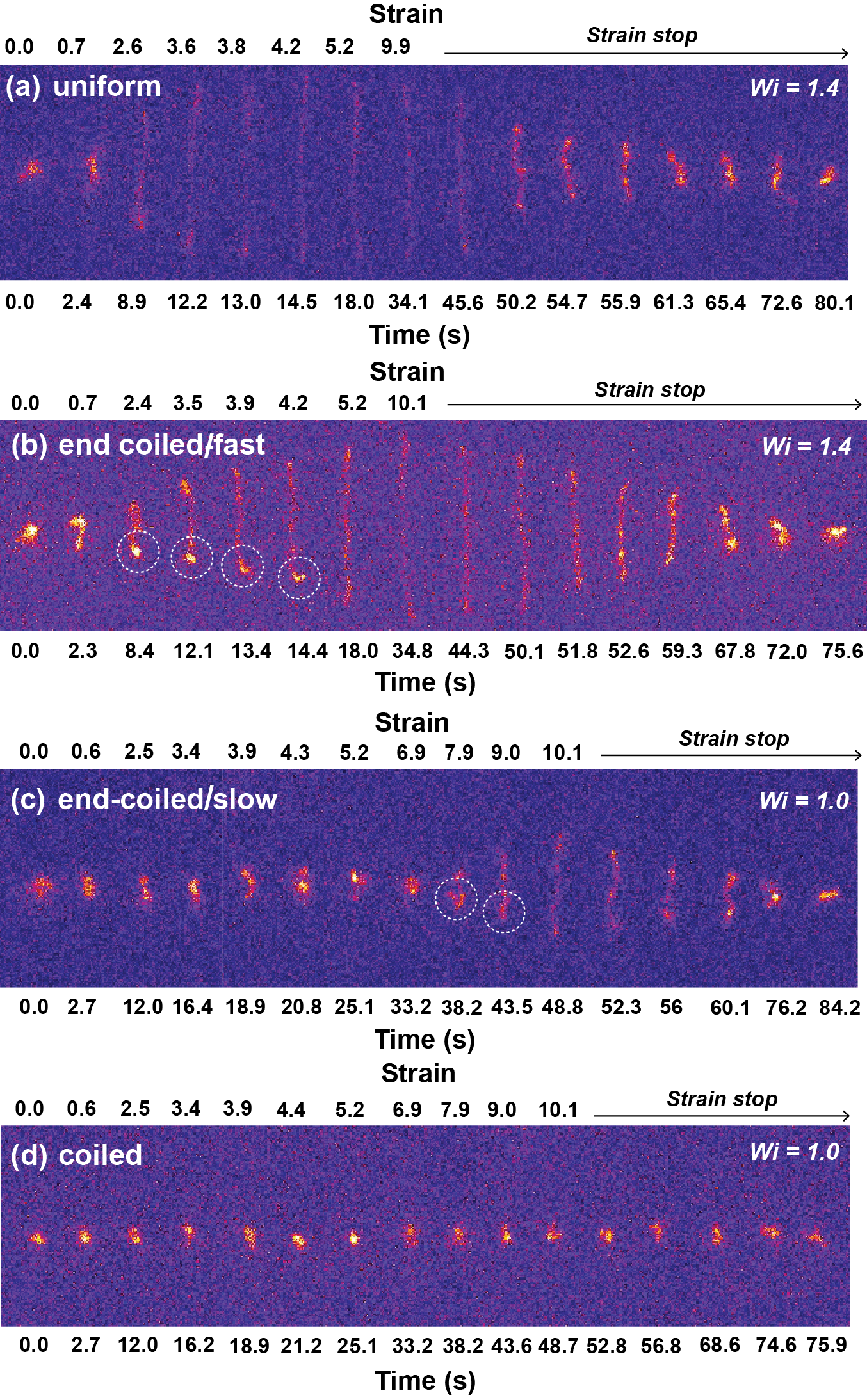}
   \caption{\label{fig:fig9}Representative single molecule images of transient polymer stretching during a step strain in extensional flow. Transient fractional extension is shown for a few different representative conformations of polymer stretch in 1 $c^{\ast}$ solutions. Molecular conformations include: (a) uniform stretch, (b) end-coiled/fast, (c) end-coiled/slow, and (d) coiled.}
\end{figure}

The broad fractional distribution arises due to a contribution of individual molecular stretching pathways that differ greatly in dynamics. Based on the single polymer stretching events, we generally observe a set of distinct molecular conformations that are classified into four categories: uniform stretch, coiled, end-coiled/fast, and end-coiled/slow (FIG.~\ref{fig:fig9}). These polymer conformations are defined using the following criteria. First, polymers in the `uniform stretch' category stretch uniformly along the contour of the backbone, with no observed kinks, folds, or visibly coiled ends. Second, `coiled' polymers remain in the coiled state throughout the deformation process. Third, polymers in the `end-coiled/fast' category show clear non-uniformity in the distribution of the backbone conformation, with one end apparently coiled during the event. Moreover, these conformations are observed to generally stretch faster than the average distribution. Finally, polymers in the `end-coiled/slow' category again show obvious non-uniformity with a clear coiled end, yet these polymers stretch slowly (more slowly than the average).

\begin{figure}[t]
\centering
\includegraphics[height=14.0cm]{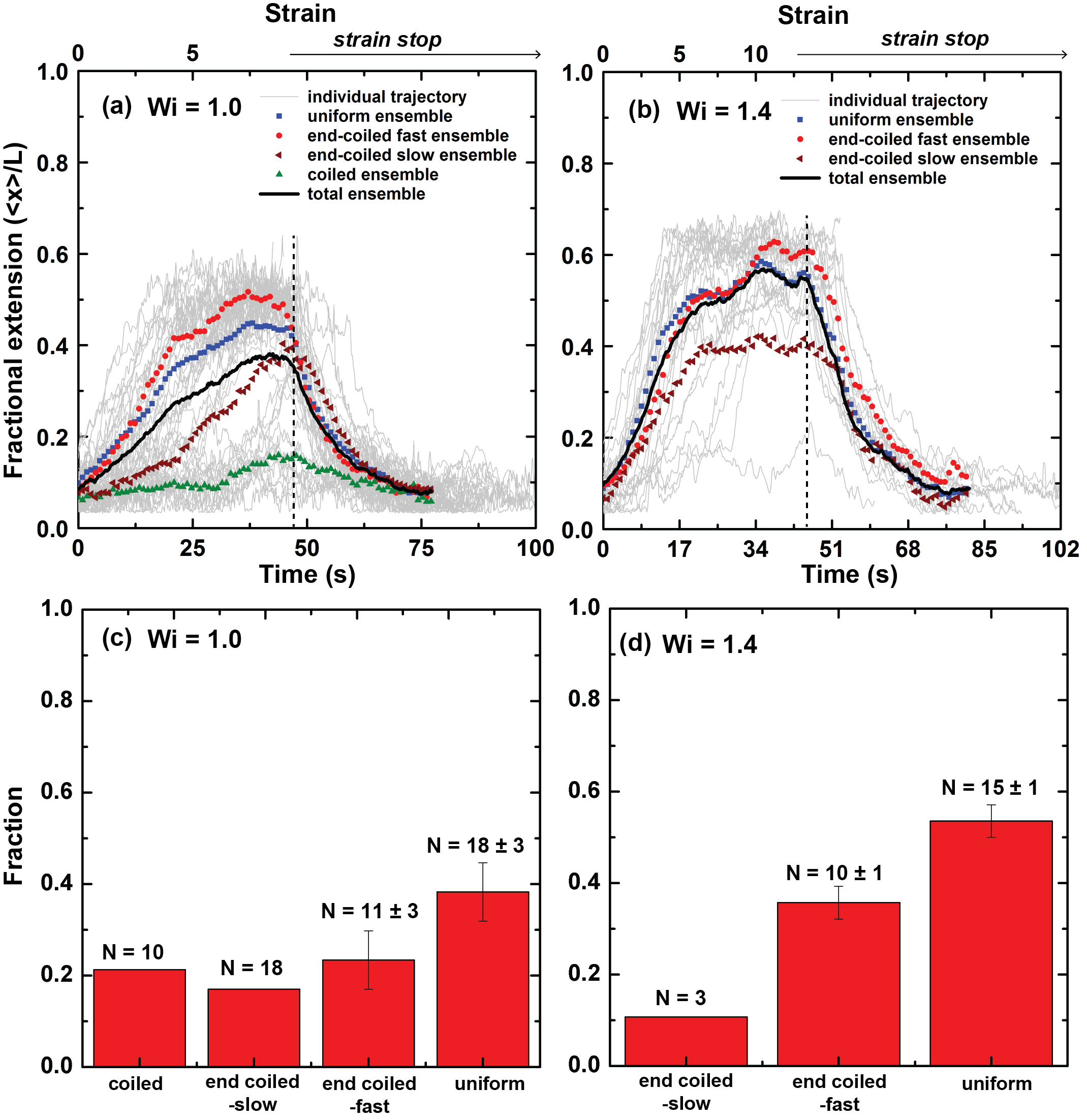}
  \caption{\label{fig:fig10}Molecular individualism in polymer stretching in semi-dilute solutions. Transient fractional extension of polymers in 1 $c^{\ast}$ solutions is shown as a function of molecular conformation. (a), (b) Transient stretching dynamics in 1 $c^{\ast}$ solutions at $Wi$ = 1.0 and $Wi$ = 1.4, with results plotted as a function of polymer conformation in terms of the ensemble average stretch. The dotted lines indicate where the step-strain rate is stopped. (c), (d) Distribution of the different molecular conformation stretching pathways at $Wi$ = 1.0 and 1.4. Error bars are determined as an uncertainty in assigning molecular conformation in a small number of trajectories.}
\end{figure}

We further analyzed the ensemble average transient stretch as a function of polymer conformation and fractional occurrence of the four confirmations (FIG.~\ref{fig:fig10}). In this way, the influence of molecular conformation on chain stretching dynamics in extensional flow is directly observed. First, we observe that polymers classified with uniform conformation generally stretch with a similar rate (or slightly faster) than the entire ensemble. Moreover, perhaps not surprisingly, polymers with coiled conformations generally do not stretch during the step strain event (FIG.~\ref{fig:fig10}a). At higher flow rates, the prevalence of coiled conformations decreases, with the complete absence of coiled molecules at $Wi$ = 1.4. Interestingly, end-coiled molecules show strikingly different dynamic behavior within the distribution for polymers with seemingly similar conformations. A sub-fraction of the end-coiled polymers stretch very slowly (end-coiled slow, FIG.~\ref{fig:fig10}a,b), whereas a different sub-fraction of end-coiled molecules (end-coiled fast) stretch quite rapidly, generally stretching at least as fast as the entire ensemble. We hypothesize that the formation of a coiled/fold structure at the terminus of the linear polymer chain facilitates the rapid stretching of a subset of the polymers classified as end-coiled, likely by formation of transient flow-induced entanglements with surrounding chains. Moreover, upon increasing accumulated fluid strain, polymers with end-coiled conformations eventually stretch out, in contrast to the fully coiled conformation (FIG.~\ref{fig:fig9} and FIG.~\ref{fig:fig10}). We conjecture that the broad distribution in transient dynamics in semi-dilute solutions arises due to intermolecular interactions.

\begin{figure}
\centering
\includegraphics[height=5.5cm]{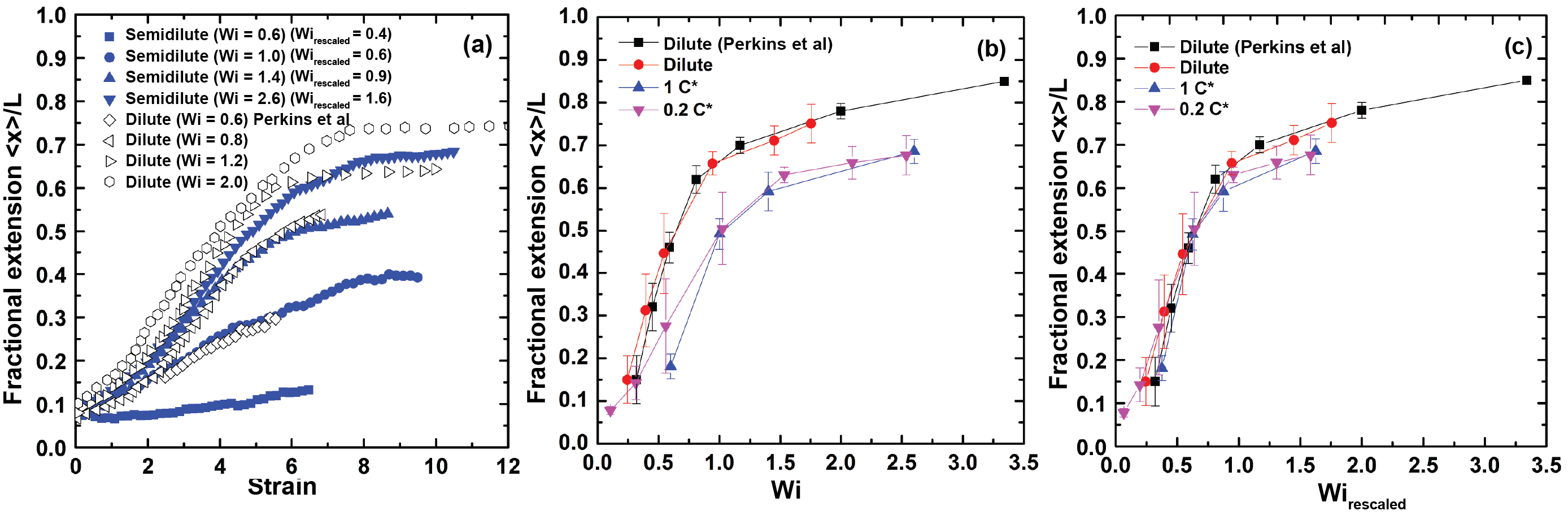}
  \caption{\label{fig:fig11}Transient and steady-state polymer stretch in dilute and semi-dilute solutions. (a) Comparison of transient fractional extension in planar extensional flow for ultra-dilute $10^{-5} c^{\ast}$ and semi-dilute 1 $c^{\ast}$ solutions at original $Wi$ and upon rescaling of $Wi$, which shows collapse of dilute and semi-dilute stretching data  (b) Steady-state fractional extension for polymers in semi-dilute (0.2 and 1.0 $c^{\ast}$) and ultra-dilute solutions ($10^{-5} c^{\ast}$) \cite{Perkins1997}. (c) Coil-stretch transition upon rescaling of $Wi$, which shows collapse of dilute and semi-dilute stretching data.}
\end{figure}

We also compared transient stretching dynamics in the start-up of planar extensional flow between dilute and semi-dilute DNA solutions (FIG.~\ref{fig:fig11}a). It is generally observed that the average transient fractional extension in semi-dilute solutions is much smaller than that in dilute polymer solutions at low $Wi$ ($Wi < 1$), where $Wi$ is defined using the longest polymer relaxation time in either dilute or semi-dilute solution. We found that the difference between transient stretch in dilute and semi-dilute solutions decreases as $Wi$ increases above 1.0, and approaches dilute transient dynamics at high $Wi$ = 2.6 (FIG.~\ref{fig:fig11}a). We also determined steady-state fractional extension for the subset of polymers that reach a steady-state extension during the experiment. In this way, the coil-stretch transition was analyzed for a polymer concentration of 0.2 $c^{\ast}$ and 1.0 $c^{\ast}$, and these data were compared to dilute solution steady-state extension data from this work and from prior literature (FIG.~\ref{fig:fig11}b) \cite{Perkins1997}. In semi-dilute solutions, a strong inhibition of chain stretching as reflected in the steady-state fractional extension is observed, with a clear difference in the coil-stretch transition between dilute and semi-dilute polymer solutions. The milder coil-stretch transition in semi-dilute solutions suggests that the critical $Wi_c$ at the coil-stretch transition may be concentration dependent. To test this hypothesis, we calculated a critical $Wi_c$ in a logarithmic scale between the coil and stretch limits, where $Wi_c$ occurs when the square fractional extension reaches the half maximum point in fractional polymer stretch, such that $ln(\bar{x}^2)= ( ln \langle \bar{x} \rangle_0^2 + ln\langle \bar{x} \rangle_{max}^2) / 2$, where $\langle \bar{x} \rangle_0 = \langle x \rangle_0 / L$ is the average near equilibrium fractional polymer extension at $Wi \approx 0$ and $\langle \bar{x} \rangle_{max} = \langle x \rangle_{max} / L$ is the average maximum fractional extension observed in our experiments far above the coil-stretch transition $Wi \gg 1$. This approach is inspired by recent work in applying Brownian dynamics simulations to study dynamic transitions in flow \cite{Cifre2001}. Using this method, we found that $Wi_c$ = 0.45 for ultra-dilute polymer solutions, whereas $Wi_c$ = 0.72 for semi-dilute solutions at both 0.2 $c^{\ast}$ and 1.0 $c^{\ast}$. We rescaled the semi-dilute $Wi$ with the ratio $Wi_{c,dilute}/ Wi_{c,semi-dilute}$, and {\em both} the transient and steady-state stretch data are found to appear to collapse between dilute and semi-dilute polymer solutions (FIG.~\ref{fig:fig11}a and FIG.~\ref{fig:fig11}c, respectively).

\section{Conclusions}
In this work, we use single molecule fluorescence microscopy to investigate the dynamics of dilute and semi-dilute DNA solutions, including relaxation from high stretch and transient and steady-state extension in extensional flow. Our results show that data on polymer relaxation in semi-dilute solutions in consistent with the scaling relations for semi-flexible polymers in the good solvent regime. Furthermore, a broad distribution of transient fractional extension in the start-up of extensional flow is observed. By comparing dilute and semi-dilute polymer transient stretching dynamics, we observe a decrease in fractional extension for the semi-dilute case compared to dilute solutions, and this difference decreases as $Wi$ increases above 1.0. We further observe fairly large differences in steady-state fractional extension between dilute and semi-dilute polymer solutions, which occurs when the $Wi$ is defined using the longest polymer relaxation in the respective dilute or semi-dilute solution. In this way, a milder coil-stretch transition for semi-dilute solutions compared to ultra-dilute solutions is observed. Indeed, the milder coil-stretch transition for semi-dilute solutions is consistent with prior flow birefringence experiments on synthetic polymers. Moreover, our experiments show a strong coupling between flow field, polymer conformation, and polymer chain-chain interactions that as a whole effect the dynamics of semi-dilute polymers in strong flows.

Our experimental results on transient stretching dynamics show that the difference in polymer stretch between dilute and semi-dilute solutions generally decreases as the $Wi$ increases. These results could suggest that individual polymer chains experience a more `dilute-like' environment at higher $Wi$. Moreover, steady-state extension data at 0.2 $c^{\ast}$ and 1.0 $c^{\ast}$ both show a milder coil-stretch transition compared to ultra-dilute solutions. Interestingly, the rescaled $Wi$ based on the ratio of critical $Wi_c$ between dilute and semi-dilute solutions leads to a collapse of both the transient and steady-state extension data. These results suggest that the critical $Wi_c$ at the coil-stretch transition is a function of polymer concentration, with different dynamic behavior observed in semi-dilute solutions. 

In prior work, a decrease in steady-state fractional extension has been conjectured to arise from some combination of polymer chain degradation, formation of transient knots or kinks along the polymer backbone, and self-entanglements. By using single molecule experiments, we are able to directly visualize individual molecules in semi-dilute polymer solutions under strong extensional flow. No evidence of polymer chain degradation or persistent kink structures along DNA backbones was observed. Therefore, we believe that the decrease in fractional extension is strongly related to intermolecular interactions that alter the local concentration and give rise to transient flow-induced entanglements. 

Our results further provide a new set of polymer conformations in semi-dilute solutions in extensional flow. In prior work, Smith et al. \cite{Smith1998} studied transient polymer stretching in the onset of extensional flow, which generally revealed polymer conformations including folds, kinks, and dumbbells. Indeed, the stretching dynamics of individual polymers in the ultra-dilute limit is highly dependent on initial molecular conformations, with the non-uniform shapes such as `folds' and `kink' leading to a slower stretch rate. In semi-dilute solutions, however, we find that the stretching rate of individual polymers depends on both the initial configurations and the surrounding background. In particular, we observe non-uniform conformations such as end-coils that lead to significant changes in polymer stretching rate: polymers with end-coil conformations either stretch much faster or much slower compared to the ensemble average. In summary, these results show clear differences between dilute and semi-dilute solutions.

\section{Acknowledgments}
We thank Prof. Prabhakar Ranganthan, Prof. Randy Ewoldt, and Dr. Folarin Latinwo for insightful discussions. We thank Dr. Christopher Brockman and Prof. Melikhan Tanyeri for implementing the hydrodynamic trap for this project. This work was funded by a Dow Graduate Fellowship for KWH and the David and Lucile Packard Foundation, NSF CAREER Award CBET-1254340, and the Camille and Henry Dreyfus Foundation for CMS.    

\bibliography{citation}

\end{document}